\documentclass[pre,10pt,aps,twocolumn]{revtex4}
\usepackage{graphicx} 
\usepackage{amsfonts} 
\usepackage{euler}

\def\s2{{\mathcal S}_2} 

\def\ts{{\bf t}(s)}
 
\def\t{{\bf t}}

\begin{document}
\title{Geometric phases and polarization patterns in multiple light
  scattering} \author{ \bf A.C.  Maggs, V. Rossetto}
\affiliation{Laboratoire de Physico-Chimie Th\'eorique, CNRS-ESPCI, 10
  Rue Vauquelin, 75005 Paris, France.}

\begin{abstract}
  Multiple light scattering is widely used to characterize dense
  colloidal systems as well as in deep tissue imaging; experiments are
  often interpreted via a theory of diffusion of the light intensity
  within a sample, neglecting the vector nature of the electromagnetic
  wave. Recent experiments on diffuse backscattering with linearly
  polarized light from colloidal suspensions of micron size particles
  were found to display strong intensity variations with fourfold
  rotational symmetry when observed through an arbitrarily oriented
  linear analyzer. We show that these polarization patterns are
  manifestations of a Berry phase of the multiple scattered beam.
\end{abstract}
\maketitle

\begin{figure}[ht]
  \includegraphics[scale=.55] {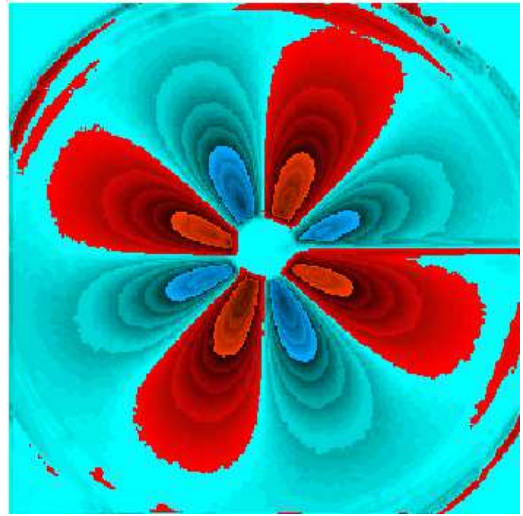}
\caption{
  False color image of polarization patterns \cite{thanks} on the
  surface of a beaker of a colloidal solution {\it \cite{mueller}}.
  Linearly polarized light is incident at a spot at the center of the
  image.  The surface of the beaker is imaged with a camera through a
  linear analyzer.  The experimentalists have plotted an element of
  the Mueller matrix characterizing the transfer function between the
  incident and backscattered light as a function of position on the
  surface.  For $2\mu m$ diameter beads the other elements of this
  matrix implied in the scattering of linearly polarized light are
  circularly symmetric; the image thus reflects the variation of
  brightness of a backscattered linearly polarized beam observed with
  a linear polarizer. Red corresponds to brighter than average
  regions, blue to darker areas. }\label{daisy}
\end{figure}

The quality of imaging in strongly scattering media such as biological
tissue is enhanced if polarization discrimination is used to filter
the light. Rather surprisingly, it is found that circularly and
linearly polarized light do not display the same quality and
resolution in imaging {\it \cite{imaging}}. These puzzling
observations have motivated a detailed characterization of the
backscattering characteristics of strongly scattering media.
Systematic experiments characterizing of the scattering properties of
dilute colloids of latex beads in solution {\it \cite{photo}} as a
model of tissue scattering found surprisingly rich results. In the
experiments a beaker of a strongly scattering suspension is
illuminated by a polarized light source focused to a small point. The
surface of the beaker is then imaged with various analyzers. With a
linear analyzer and a linearly polarized beam the experiments on $2\mu
m$ diameter latex beads showed strong variations in intensity
exhibiting a fourfold symmetry about the incident spot, somewhat like
the petals of a daisy, Fig.  (\ref{daisy}).  An additional striking
result is that when the analyzer is rotated an angle $\pi/4$ the
pattern of intensity rotates just one half this angle, $\pi/8$,
without changing shape.  While analytic approaches {\it \cite{weitz}}
have been successful in treating problems in the propagation of the
intensity of multiple scattered light, many treatments of the
evolution of the polarization state have been purely numerical, {\it
  \cite{monte,these,maynard}} using Monte-Carlo techniques to trace
light through a multiply scattering medium.  Analytical work has been
based on the idea that polarization states should be rapidly
randomized and that polarization dependent effects should be weak and
transient {\it \cite{stanford}}.  We show in this report that simple
geometric considerations allow one to gain a qualitative understanding
of the observed polarization patterns. The patterns are due to Berry
phases in the multiply scattered beam.

The evolution of the plane of polarization of light propagating in a
{\sl smoothly} disordered medium was first treated by Rytov {\it
  \cite{r1}} in the eikonal approximation. The geometry of the
propagation was then rediscovered by Berry {\it \cite{berry}} and
applied to many other wave phenomena, including quantum mechanics. The
results of these studies are best illustrated by experiments in which
light propagates along a tortuous fiber optic {\it
  \cite{optics,haldane}}. While propagating along a uniform fiber the
polarization state of the light evolves in such a way to minimize
twisting {\it \cite{acm}} of the polarization vector ${\bf E}$, in
fact one speaks of the evolution of ${\bf E}$ by {\sl parallel
  transport}. This evolution law is locally trivial, however it leads
to a global rotation of the polarization state. This effect of
coupling of polarization to the propagation path is now generally
considered as a simple example of a {\sl geometric phase}.  The main
conclusion of the present letter is that identical phenomena are to
expected even in the absence of the guiding fiber, as for instance is
the case for light multiply scattered in a colloidal suspension; In
this report we shall firstly justify the use of the Rytov-Berry
result, valid for the evolution of the polarization in a continuous
medium, in the case of multiple scattering by distinct particles.  We
then show that a geometric phase naturally leads to fourfold symmetric
polarization patterns in retro-diffusion.


If the polarization state of a light beam is to evolve by parallel
transport we require that helicity flipping events are rare; such
events are for instance generated by large angle deflections such as
reflections at an interface. Specular reflection at a surface
preserves the linear polarization state of an incident beam and can
not lead to fourfold symmetric patterns in the back scattered beam. It
has been shown both numerically {\it \cite{these,maynard}} and
analytically {\it \cite{goro}} that helicity flipping events occur on
a characteristic length scale which is somewhat larger than the length
over which the the beam is deviated in the case of strongly forward
scattering. From now on we neglect these events. Furthermore using the
Born approximation it has been shown {\it \cite{mackintosh,goro}} that
under multiple scattering conditions the polarization vector of the
forward scattered beam evolves so that
\begin{math}
  {\bf E }_j \sim {\bf E}_{j-1}- ({\bf E}_{j-1}.  {\bf t}_j) {\bf t}_j
\end{math},
where ${\bf E }_j $ is the polarization vector after the $j$'th
collision and ${\bf t}_j$ is the direction of propagation of the
light.  In the limit of many small angle scattering events this
evolution law is equivalent to parallel transport of the polarization
vector.  This result is valid for scattering from micron sized latex
particles, used in the experiments.

\begin{figure}[ht]
  \includegraphics[scale=.45] {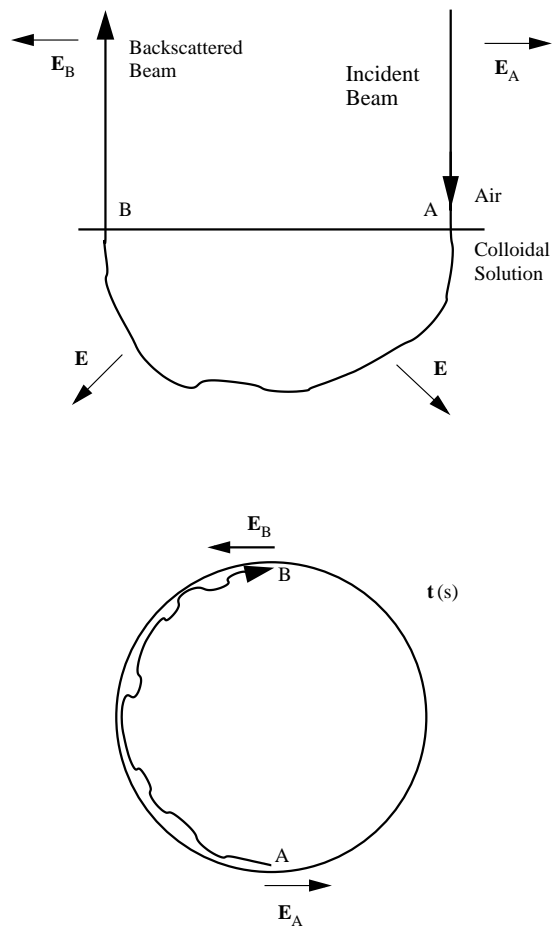}
\caption{
  { {\sl top:} Light is incident on a multiple scattering medium. The
    direction of propagation is randomly bent back and the light
    escapes from the surface.  {\sl bottom} } The direction of
  propagation of the scattered light is plotted on a sphere. Incident
  light corresponds to the south pole, $A$. The escaping light
  corresponds to the north pole, $B$. The indicated path is scattered
  principally to the left so that on the sphere the path remains, on
  average on the western area of the sphere. As the point of
  observation goes around the incident beam, an angle of $2\pi$, the
  path between the poles sweeps out an area on the sphere of $4\pi$.
  Polarized light with polarization state ${\bf E}_A$ is parallel
  transported to the state ${\bf E}_B$.  }\label{backscatter}
\end{figure}

The angle of rotation of a polarization vector due to a geometric
phase is calculated from the propagation direction ${\bf t}(s)$
expressed as a function of the path length $s$.  It is identical to
the solid angle enclosed by the path $\ts$ on a sphere {\it
  \cite{berry,haldane}}.  We now proceed by translating the
backscattering geometry into an ensemble of paths on the unit sphere,
$\{\t \}$ in order to apply this result.  As shown in Fig.
(\ref{backscatter}) backscattered light corresponds to a path from the
south to north poles of the sphere, $\ts$, describing the direction of
propagation.  We take as a reference state light scattered to the
left, polarized in the plane of the page. We see that an original
polarization vector ${\bf E}_A$ is parallel transported around the
sphere, Fig.  (\ref{backscatter}, bottom) so that the initial and
final vectors are antiparallel. The indicated path Fig
(\ref{backscatter}, top) is scattered preferentially to the left;
since the real space scattering of the beam is linked to the tangent
curve by the equation
\begin{math}
  {\bf r}(s) = \int^s_0 {\bf t}(s') \; ds',
\end{math}
the path $\ts$ must remain largely on the western latitude as shown in
the Fig (\ref{backscatter}, bottom).  We now change the point of
observation on the surface of the sample, generating a second
trajectory ${\bf t'}(s)$.  This new trajectory together with the
reference path form a closed loop on the sphere which allow us to
apply the Berry result.  As we change the point of observation on the
sample, $B$, and wind an angle of $2\pi$ about the incident beam, $A$,
the new path $ {\bf t'}(s) $ on the sphere sweeps out a solid angle of
$4 \pi$.  We thus deduce that the polarization at the surface of the
sample rotates {\sl two full turns}\/ as we move just once about the
incident beam.  Since a linear analyzer is sensitive to the angle of
rotation modulo $\pi$ we understand that there are four radial
directions in which an analyzer detects a maximum in the intensity. We
also understand that if the analyzer rotates an angle $\theta$ then
the intensity pattern rotates just $\theta/2$.

The geometry of this result is strongly reminiscent of the {\sl plate
  trick} demonstration for spinor rotations {\it \cite{spinor}}.  If
one holds a plate horizontally in the palm of one's hand one can spin
it about a vertical axis by performing a suitable contortion of the
arm.  Against all intuition the plate can be turned an arbitrarily
large angle; for each single cycle of the arm the plate spins twice.
Parameterizing the shape of the arm by its direction $\ts$ and
plotting this on the sphere one sees that this trick can also be
understood by the fact that during a cycle $\ts$ sweeps out a solid
angle of $4 \pi$ on a sphere translating giving two full rotations in
a plane for each cycle.

Clearly the geometry of Fig. ({\ref{backscatter}}) is simplified; we
have neglected fluctuations in the light paths. In order to calculate
the full illumination state of the light at $B$ one should sum over
all scattering paths from $A$ to $B$; the path sketched in Fig.
(\ref{backscatter}) is just one contribution to this sum. However,
paths that exit at the same point of the surface are correlated so
that the Berry phases corresponding to each path are also correlated
and do not average to zero.  The superposition leads to modifications
in the geometry when imaging at small distances from the incident
beam. For light which returns directly on axis, all paths around the
sphere are equally likely and the polarization state is undetermined
{\it \cite{mackintosh}}.  Away from the central spot the sum is
dominated by the most direct paths, such as the path shown in the
figure.  We note that with a coherent light source the speckle
structure of the radiation field can never be neglected. There are
naturally strong fluctuations in the polarization state due to the
existence of zeros in the instantaneous speckle pattern for each
component of the polarization vector. The experimental image, Fig.
(1), is in fact a time average so no speckles can be seen.

In this report we have considered the problem of scattering from
particles in the Mie and Rayleigh-Gans scattering regimes in the limit
of strong forward scattering where individual scattering events show
weak polarization dependence.  Small particles have very different
scattering properties and show strong polarization dependence in the
scattering.  In this limit {\it \cite{monte}} the fourfold symmetric
pattern due to parallel transport is not seen.  Our results can also
be used to understand the scattering of circularly polarized light. In
this case the Berry phase is just a simple phase shift of the
backscattered beam, rather than a rotation in polarization plane. No
polarization patterns are to be expected on the surface of a uniform
colloid. We understand that the difference in the coupling of the
Berry phase to linearly and circularly polarized light is partially
responsible for the different imaging qualities of circularly
polarized light in colloidal suspensions and tissue fantoms {\it \cite
  {bio} }. An object hidden deep under the surface of the beaker in
Fig. (1) can only weakly modify the intensity at the surface. This
weak modification is easily hidden by the strong variation due to the
Berry phase.

Finally, a number of interesting variations can be played on the
geometry of the figures observed in diffuse backscattering.  Addition
of an strongly optically active molecule in the solution should lead
to systematic bending of the daisy giving rise to helical patterns in
the polarization state on the surface of the colloidal solution. This
would be of interest as a method of measuring path lengths in the
solution as a function of the distance between the incident beam and
the detection point.


\vskip 0cm
\begin{thebibliography}{99}
\bibitem{imaging} {\it Opt.-OT}. {\bf 38}, 4252-4261 (1999)

  
\bibitem{photo} {\it Proceedings-of-the-SPIE} {\bf 2976}, 298-305
  (1997).
   
\bibitem {weitz} D.A.~Weitz, D.J.~Pine, in {\it Dynamic Light
    Scattering: The Method and some Applications} pp. 652-720 W.
  Brown Ed.  (Clarendon Press, Oxford, 1993).
    
\bibitem{monte} S.~Bartel, A.H.~Hielscher, {\it Appl. Opt.} {\bf 39},
  1580-1588 (2000).
 
\bibitem{these} A.S.~Martinez, Thesis, Univ.  Joseph Fourier, Grenoble
  (1993).
 
\bibitem{maynard} A.S.~Martinez, R.~Maynard, in {\it Localization and
    Propagation of Classical Waves in Random and Periodic
    Structures}\/ C.M.  Soukoulis Ed. 99-114 (Plenum Publishing Corp.
  N.Y.  1994).
  
\bibitem{stanford} M.~Moscoso, J.B.~Keller, G.~Papanicolaou, {\it J.
    Opt. Soc. of Am.}  {\bf A18}, 948-960 (2001).
    
\bibitem {r1} S.M.~Rytov, {\it Dokl. Akad. Nauk.  USSR}, {\bf 18}, 263
  (1938).  Reprinted in {\sl Topological phases in quantum theory}
  (World Scientific, 1989).
  
\bibitem {berry} M.V.~Berry, {\it Nature}, {\bf 326}, 277-278 (1987).
 
\bibitem {optics} A.~Tomita, R.Y.~Chiao, {\it Phys. Rev.  Lett.}  {\bf
    57}, 937-940 (1986).
 
\bibitem{haldane} F.D.M.~Haldane, {\it Opt. Lett.} {\bf 11}, 730-732
  (1986).
    
\bibitem{acm} A.C.~Maggs {\it J. Chem. Phys.} {\bf 114}, 5888 (2001).
  
\bibitem{goro} E.E.~Gorodnichev, A.I.~Kuzovlev, D.B.~Rogozkin, {\it
    JETP lett.}  {\bf 68}, 22-28 (1998).
    
\bibitem{mackintosh} F.C.~Mackintosh, J.X.~Zhu, D.J.~Pine, D.A.~Weitz,
  {\it Phys.  Rev.  }  {\bf B40}, 9342-9345 (1989).
    
\bibitem{spinor} R.P. Feynman, in {\it Elementary particles and the
    laws of physics : the 1986 Dirac memorial lectures.}\/ (Cambridge
  University Press, 1987).
    
\bibitem{mueller} A.H.~Hielscher {\it et al.}  {\it Optics exp.}  {\bf
    1}, 441-453 (1997).

  
\bibitem{bio} V.~Sankaran, M.J.~Everett, D.J.~Maitland, J.T.~Walsh Jr.
  {\it Opt.  Lett.} {\bf 24}, 1044 (1999).
  
\bibitem{thanks} We would like to thank A.H.~Hielscher for permission
  to use Fig.  (1).

\end{thebibliography}
\end{document}